# Performance Evaluation of Spectrum Mobility in Multi-homed Mobile IPv6 Cognitive Radio Cellular Networks

Mehran Shoushtari Moghaddam, Naser Movahhedinia, Mohammad-Reza Khayyambashi
Faculty of Computer Engineering
University of Isfahan
Isfahan, Iran
E-mail: {mhm13819, naserm, m.r.khayyambashi}@eng.ui.ac.ir

Faramarz Hendessi
Department of Electrical and Computer Engineering
Isfahan University of Technology
Isfahan, Iran
E-mail: hendessi@cc.iut.ac.ir

*Abstract*— **Technological developments alongside VLSI achievements enable mobile devices to be equipped with multiple radio interfaces which is known as multihoming. On the other hand, the combination of various wireless access technologies, known as Next Generation Wireless Networks (NGWNs) has been introduced to provide continuous connection to mobile devices in any time and location. Cognitive radio networks as a part of NGWNs aroused to overcome spectrum inefficiency and spectrum scarcity issues. In order to provide seamless and ubiquitous connection across heterogeneous wireless access networks in the context of cognitive radio networks, utilizing Mobile IPv6 is beneficial. In this paper, a mobile device equipped with two radio interfaces is considered in order to evaluate performance of spectrum handover in terms of handover latency. The analytical results show that the proposed model can achieve better performance compared to other related mobility management protocols mainly in terms of handover latency.**

## I. Introduction

Recent innovations in wireless networking systems and communication technologies, allow mobile devices to be equipped with multiple radio interfaces; which are called multi-homed mobile devices, and maintain their connections while moving across various wireless access technologies. The integration of heterogeneous wireless access technologies have led to the trend toward Next Generation Wireless Networks (NGWN). Multi-homing deployed in NGWN has many advantages like: resilience, load balancing and ubiquitous access support [1]. One of the main concerns in NGWN is providing seamless and ubiquitous connections to mobile devices while their point of attachment to the network changes. Mobile IPv6 as a solution toward mobility management in NGWN has been introduced to allow a Mobile Node (MN) to maintain its connection to the network regardless of its location [2]. Handover is the process by which a MN keeps its connection active while moving across one point of attachment to another. During this process, the MN cannot send or receive any data packets because of link and network layers operations delay. Vertical handover issues have been studied in the literature [3], [4], [5]. To decrease handoff latency some extensions of MIPv6 such as Hierarchical Mobile IPv6 (HMIPv6), Fast handover for Mobile IPv6 (FMIPv6), Fast handover for Hierarchical Mobile IPv6 (FHMIPv6) and Proxy Mobile IPv6 (PMIPv6) have been proposed by Internet Engineering Task Force (IETF) [6], [7], [8], [9], [10].

Cognitive Radio Networks (CRNs) were introduced as a promising solution to overcome wireless bandwidth utilization inefficiency which is deployed in NGWNs [11], [12], [13], [14]. The fundamental idea of CRNs is cooperation of two types of users: primary users (PUs) and secondary users (SUs). When a PU requests the radio spectrum resources, the SU switches to another free spectrum accordingly in order to not interfere with the transmission of the PU. This process is referred to as spectrum mobility or spectrum handover, which is the unique mobility characteristic in cognitive radio cellular networks. This feature is sophisticated by spectrum sensing functionality in the cognitive radio device [12].

In [15] a novel analytical model is developed for comparison of various mobility management protocols in terms of handover latency, as well as packet density, and packet arrival rate during the handover time. Constructing a new Care of Address (nCoA) and performing Duplicate Address Detection (DAD) in the New Access Router (NAR) in advance to reduce handover latency, and sending Binding Update (BU) to the Home Agent (HA) and Correspondent Node (CN) through Previous Access Router (PAR) to reduce the registration latency on the other hand, is the approach proposed in [16] which is known as Enhanced Fast handover for MIPv6 (E-FMIPv6). The idea of multi-homed fast handoff scheme has been proposed in [17], [18]. Based on L2 triggers and multi-homed techniques authors acclaimed handoff latency and packet loss have been reduced. However, in classical multi-cell based networks, L2 handover issues are the major concern.

None of the approaches mentioned above consider the fluctuating nature of radio spectrum resources in CRNs to estimate handover latency.

On the other hand, several cross layer handoff management techniques in NGWN have been proposed in the literature [19]. A cross layer protocol of spectrum mobility and handover in cognitive LTE Networks with the consideration of the minimum

*This research is supported by ITRC (Iranian Telecom. Research Centre).*

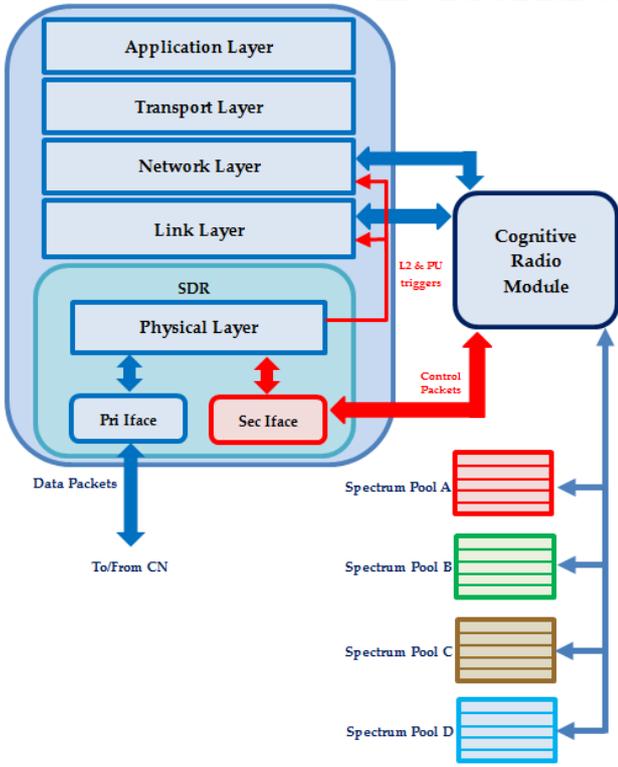

Fig. 1. The Mobile Cognitive Radio User architecture.

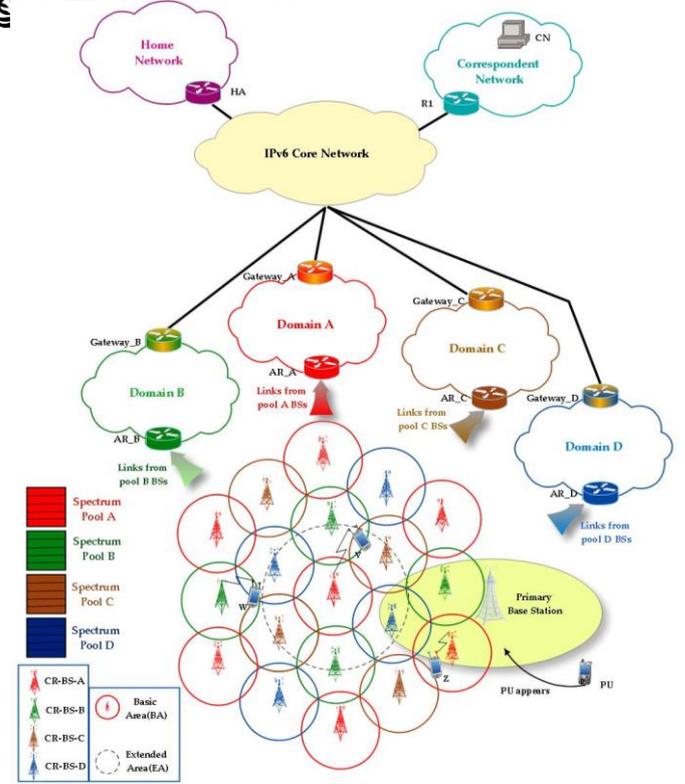

Fig. 2. Proposed system model.

expected transmission time has been proposed in [20]. In [21] authors proposed a mobility management framework in cognitive radio cellular networks. Two aspects of mobility management, spectrum mobility and user mobility management have been investigated. None of the above mentioned methods consider the multi-homed mobile device in order to evaluate mobility management.

Using MIPv6 protocol in order to manage the mobility of users, is more beneficial than Mobile IPv4 in terms of signaling overhead, handover latency, inherent security and so on [6]. So, this research investigates managing multihomed mobile users in the context of cognitive radio networks to improve the handover process in terms of handover latency. Providing fast, smooth and seamless handover in various mobility events alongside provisioning acceptable quality of service while considering heterogeneous access networks is among contributions.

The remainder of this paper is organized as follows: Section II offers the proposed system architecture. Section III describes the spectrum mobility for a multi-homed MN. After that the analytical model is presented in section IV in order to evaluate the handover latency. Finally numerical results from the performance evaluation perspective are investigated in section V before concluding remarks drawn in section VI.

## II. BASIC SYSTEM MODEL

In this work an analytical model for evaluating the latency of various MIPv6 handoff types in cognitive radio cellular networks has been developed. In the literature SUs are called, Mobile Cognitive Radio users (MCR user), which are equipped with primary radio interface and the secondary one. The former is used for regular data transmissions while the latter is used for management and control purposes. The overall architecture of MCR user has been shown in Fig. 1. When a PU appears in the spectrum or L2 triggers such as continuous reduction of RSS, increase of BER, changing the QoS and etc. are arisen, a notification to the upper layer (network layer) is sent through the secondary radio interface. Upon receipt of event notification in the network layer the role of two radio interfaces is swapped; the formerly secondary radio interface becomes primary and used for data communication purposes and the formerly primary radio interface becomes secondary used for control and management tasks.

The spectrum pooling concept which is deployed in [21] is used here. Each spectrum pool consists of several spectrum bands. Furthermore each cell consists of two coverage areas; *Basic Area (BA)* and *Extended Area (EA)*. The BA covers the current cell not overlapped with neighbor cells and consists of multiple spectrum bands; known as basic spectrum bands, whereas the EA overlaps with neighbor cells and consists of single spectrum band; known as *Extended Spectrum* (ES) band. It is assumed that one spectrum pool is assigned to each cell independently of the other cells. The cell which has the same spectrum pool as the current cell is referred to as *Extended Neighbor* (EN). Besides it is assumed that each cell with identical spectrum pool characteristics belongs to the same AR. According to PU appearance, various handoff types arise. The *Intracell/Intrapool* handoff occurs when MCR user switches to another spectrum band in the same spectrum pool of current cognitive radio base station (CR-BS). The *Intercell/Interpool* handoff occurs when MCR user switches to another CR-BS which has different spectrum pool with the current CR-BS. The MCR user has to reconfigure its RF front end in the Intercell/Interpool handoff. Fig. 2 shows the overall proposed system architecture used thorough the entire paper.

In this paper one aspect of mobility management, spectrum mobility is investigated, which is happened whether a PU appears in the spectrum band or current spectrum conditions become worse such as cell overload.

## III. SPECTRUM MOBILITY

In the spectrum mobility management point of view, two situations are considered; 1) MCR user presents in the BA and 2) MCR user presents in the EA as shown in Fig. 3. First, while MCR user is in the BA of current cell, the primary radio interface is communicating with the CN, whereas the secondary one proactively sensing radio environment and performing link layer related measurement operations such as RSS monitoring, BER estimation and etc. In the proposed method the MCR user proactively evacuates the channel prior to the PU appearance using the predicted information obtained by past channel histories and spectrum sensing capability. Fig. 4 shows the proactive spectrum handover method. Spectrum mobility management functionality has three cases to be considered which is derived in the following subsections.

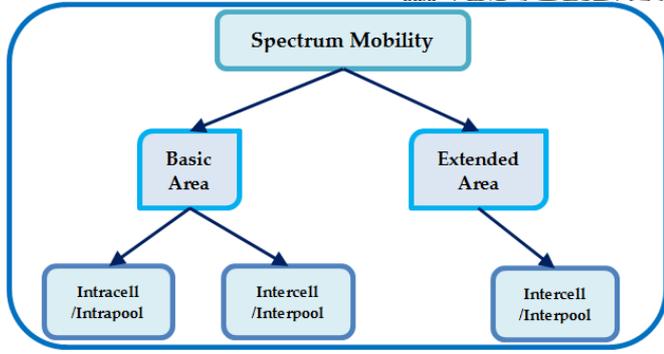

Fig. 3. Spectrum mobility management hierarchy.

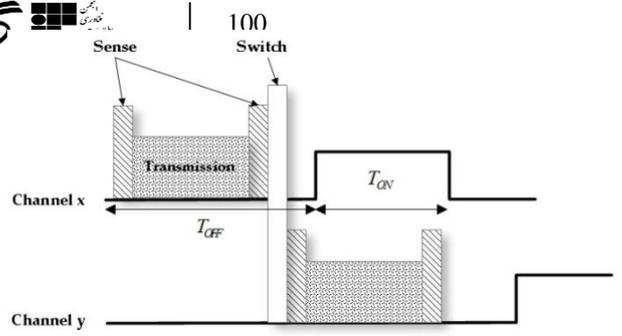

Fig. 4. Proactive spectrum handoff mode.

### A. Case A

Once the PU arrival is predicted by the secondary radio interface, it informs the upper layers about PU appearance and switches to another spectrum hole if there are available spectrum bands in the current cell. The handoff type in this case is Intracell/Intrapool handoff. In the meantime the primary radio interface is transceiving data packets, the secondary radio interface is performing spectrum sensing and link layer functionalities according to past channel histories. Using the proactive spectrum handoff, the probability that a specific spectrum is idle in the next time slot is predicted. Using the observation and predicted results the secondary radio interface switches to another free spectrum band intelligently according to spectrum decision function. After spectrum mobility, the role of two interfaces is swapped as explained before. In this case the network prefix advertised by the current AR does not change; therefore the MIPv6 handover latency is equals to interface switching which is approximately zero.

### B. Case B

If there are not available spectrum bands in the current cell due to PU activity or cell overload, the MCR user should relinquish the current cell and move to the neighbor cell having different spectrum pool. The secondary radio interface reconfigures its RF front end and switches to another available spectrum band. As the previous case the role of two radio interfaces are swapped. The handoff type in this case is Intercell/Interpool handoff.

### C. Case C

If PU activity is detected when MCR user is in the EA of the current cell, it should perform Intercell/Interpool handoff because it cannot find any available band in that area. Similar to previous cases, the role of two interfaces are swapped when PU activity is detected in the EA. Because the network layer parameters such as network prefix changes in the last two cases (cases B and C), i.e. the AR is changed while MCR user is handed over to the neighbor cell, it is desirable to use the enhanced fast handover scheme proposed in [14] to reduce MIPv6 handoff latency. When it is determined that Intercell/Interpool handoff is about to happen, first the appropriate cell is selected which the MCR user is going to move to through stochastic connectivity factors. Then the secondary radio interface reconfigures its RF front end and the following steps are performed [16]:

*Step 1:* The MCR user requests nCoA from NAR through PAR. Accordingly PAR sends out the related information of MCR user to NAR. Then the NAR generates a new CoA and performs DAD. The newly generated CoA is advertised to MCR user through PAR. At the same time the BU procedure with the HA and CN is performed by PAR.

*Step 2:* After the MCR user moves to the neighbor cell, it will send a router solicitation with the Fast Neighbor Advertisement (FNA) option. Then NAR will respond with a router advertisement with the FNA_Ack option.

## IV. ANALYTICAL MODEL

In this section the models for various handoff components delays are derived for analyzing the performance of aforementioned proposed system.

The PU arrival is modeled by Poisson process with the average arrival rate of $\lambda$, so the PU inter-arrival time $t_{pu}$ follows an exponential distribution. Hence the average idle (OFF) period of PU is $1/\lambda$. Similarly the length of busy (ON) period known as PU call holding time $t_{cp}$ follows an exponential distribution. According to the PU channel usage, the steady state probabilities of OFF state, $P_{off}$ and ON state, $P_{on}$ can be expressed as follows:

$$P_{off} = \frac{\mu_{cp}}{\mu_{cp} + \lambda}$$
$$P_{on} = \frac{\lambda}{\mu_{cp} + \lambda} \quad (1)$$

and the PU traffic intensity, $\delta$, is given by:

$$\delta = \frac{\lambda}{\mu_{cp}}. \quad (2)$$

The following analysis is taken from [22]. Let $N$ denote the number of channels in each spectrum band. The number of channels occupied in each spectrum band is defined as the system state. Let $i$ ($0 \le i \le N$) denotes system state, then the steady state probability of each state, $\pi_i$, according to Erlang-B formula is as follows:

$$\pi_i = \frac{\delta^i/i!}{\sum_{i=0}^{N} \delta^i/i!}; 0 \le i \le N. \quad (3)$$

The blocking probability from PU perspective, $P_b$, is when all of the channels in the spectrum band are occupied and is given by:

$$P_b = \pi_N = \frac{\delta^N/N!}{\sum_{i=0}^{N} \delta^i/i!}. \quad (4)$$

Therefore the probability that cell $i$ is not overloaded, $P_{under}^i$ and the cell overload probability, $P_{over}^i$ is given by:

$$P_{over}^i = P_b$$
$$P_{under}^i = 1 - P_b. \quad (5)$$

Let $P_L$ denotes the probability that a MCR user vacates its channel when PU appears and reclaims the channel. From the PU perspective, the probability of particular channel reclaimed by PU if the system is in state i is obtained by $1/((N-i))$. Therefore

| Symbol | Description |
|---|---|
| $h_{C-H}$ | Avg number of hops between CN and HA |
| $h_{C-G}$ | Avg number of hops between CN and Gateway |
| $h_{H-G}$ | Avg number of hops between HA and Gateway |
| $h_{G-A}$ | Avg number of hops between Gateway and AR |
| $h_{A-A}$ | Avg number of hops between neighbor ARs |
| $h_{A-BS}$ | Avg number of hops between AR and BS |
| $L_{RS}$ | Size of RS message, 52 Bytes |
| $L_{RA}$ | Size of RA message, 80 Bytes |
| $L_{BU-HA}$ | Size of BU sent from MN to HA, 56 Bytes |
| $L_{BA-HA}$ | Size of BA, 56 Bytes |
| $L_{BU-CN}$ | Size of BU sent from MN to CN, 66 Bytes |
| $L_{HoTI}$ | Size of HoTI message, 64 Bytes |
| $L_{CoTI}$ | Size of CoTI message, 64 Bytes |
| $L_{HoT}$ | Size of HoT message, 74 Bytes |
| $L_{CoT}$ | Size of CoT message, 74 Bytes |
| $L_D$ | Size of data packet 120 Bytes |

$P_L$ is the probability that one particular channel is reclaimed by PU, if there are free channels from the PU point of view.

$$P_L = Pr(\text{chan reclaim} \mid \text{available free chan}) = \frac{\sum_{i=0}^{N-1} \frac{1}{N-i} \pi_i}{P_{under}^i}. \quad (6)$$

In other words, $P_L$ is the probability that the reclaimed channel is the one that MCR user has occupied. Accordingly the probability that MCR user does not need to evacuate its channel, $P_{NL}$ is defined as:

$$P_{NL} = \frac{\sum_{i=0}^{N-1}\left(1 - \frac{1}{N-i}\right)\pi_i}{P_{under}^i}. \quad (7)$$

After vacating the channel reclaimed by PU, if there are available channels in the system the spectrum handoff is successful. Otherwise the spectrum handoff is failed. Let $p_{succ}^{sm}$ and $p_{fail}^{sm}$ denote the probability of spectrum handoff is successful and failed respectively. Their expressions are as follows:

$$p_{succ}^{sm} = P_L P_{under}^i$$
$$p_{fail}^{sm} = P_L P_{over}^i. \quad (8)$$

Let $t_{mcr}$ the MCR user service time random variable with the mean $1/\mu_{mcr}$, pdf of $f_{t_{mcr}}(t)$, Cumulative Distribution Function (CDF) of $F_{t_{mcr}}(t)$, Laplace Transform of pdf $\mathcal{L}_{f_{t_{mcr}}}(s)$ and Complementary Cumulative Distribution Function (CCDF) $\bar{F}_{t_{mcr}}(t)$, i.e. $\bar{F}_{t_{mcr}}(t) = 1 - F_{t_{mcr}}(t)$. For the sake of simplicity the exponential distribution is assumed for the MCR user service holding time. Let $H$ denotes the discrete random variable of number of spectrum handoffs. The probability of zero spectrum handoff is given by:

$$Pr(H = 0) = \mathcal{L}_{f_{t_{mcr}}}\left(\lambda(1 - P_{NL})\right). \quad (9)$$

If the MCR user can complete its service before PU appearance or the PU reclaims another spectrum band during MCR user service time, there will be no need to perform spectrum handoff. The probability for $k$ spectrum handoffs is derived as follows:

$$Pr(H = k) = \frac{\left(-\lambda p_{succ}^{sm}\right)^k}{k!} \mathcal{L}_{f_{t_{mcr}}}^{(k)}\left(\lambda(1 - P_{NL})\right) + \frac{\left(-\lambda p_{succ}^{sm}\right)^{k-1} \lambda p_{fail}^{sm}}{(k-1)!} \mathcal{L}_{f_{t_{mcr}}}^{(k-1)}\left(\lambda(1 - P_{NL})\right), \quad (10)$$

where $U^{(k)}(t)$ is the $k^{th}$ derivative of function $U$. The first term in (10) represents all of $k$ spectrum handoffs are successful and the second one represents the $1^{st}, 2^{nd}, \ldots, (k-1)^{th}$ spectrum handoff are successful and the $k^{th}$ spectrum handoff is failed. Therefore the average number of spectrum handoffs is as follows:

$$E(H) = \frac{\left(p_{succ}^{sm} + p_{fail}^{sm}\right)\left(1 - \mathcal{L}_{f_{t_{mcr}}}\left(\lambda p_{fail}^{sm}\right)\right)}{p_{fail}^{sm}}. \quad (11)$$

### A. Packet Transportation Delay over Wireless and Wired Link

Because of intrinsic erroneous feature of wireless links, unreliability is more common in wireless networks than in the wired ones. During transmission of data, some of the frames in transmit may be corrupted so the retransmission of frames being in error is inevitable in wireless communications. Hence the one way packet transportation delay over the wireless link, $d_{wl}(L_p)$ is obtained by [22]:

$$d_{wl}(L_p) = d_{fr} + (w - 1)\zeta. \quad (12)$$

where $L_p$, $d_{fr}$, $w$ and $\zeta$ are packet length, frame transportation delay, number of frames per packet and interframe time respectively.

Similarly the one way packet transportation delay over the wired link would be obtained as follows:

$$d_{wd} = \frac{L_p \times h}{BW} + D_{wd}, \quad (13)$$

where $BW$ is the wired link bandwidth, $D_{wd}$ is the wired link propagation delay and $h$ is the number of hops between source and destination.

### B. Mobile IPv6 Handover Analysis

The MIPv6 handover latency has several components. The first part includes the link layer (L2) handoff latency, $t_{L2}$ and occurs when physical and link layer parameters are changed. Then the movement detection procedure is performed to ensure changing of point of attachment, $t_{MD}$. After that, CoA configuration and Duplicate Address Detection (DAD) is performed to generate an unique CoA, $t_{DAD}$. Finally the BU registration process is done, $t_{REG}$. The notations used in this paper are summarized in Table 1 [23].

From Table 1 the average number of hops between HA and AR, $h_{H-A}$ would be $h_{H-G} + h_{G-A}$. In addition it is obvious that the number of hops between MCR user and its CR-BS is one. It is assumed that $h_{A-A}$ can be rewritten by $\sqrt{h_{G-A}}$ [23].

The expression for basic MIPv6 handover latency with a single radio interface, $L_{MIPv6}$ is derived by:

$$L_{MIPv6} = t_{L2} + t_{SM} + t_{MD} + t_{DAD} + t_{REG}, \quad (14)$$

where $t_{SM}$ is the spectrum mobility delay. In the case of basic MIPv6 with single radio interface the expression for spectrum mobility delay can be obtained by [20]:

$$t_{SM} = t_{prep} + t_{rcfg} + t_{syn}^{sen} + t_{sen} + t_{dec} + t_{syn}^{tx}, \quad (15)$$

TABLE 2
System Value Parameters

| Parameters | Symbols | Values |
|---|---|---|
| Inter-frame time | $\zeta$ | 30 ms |
| Number of retransmission trials | $n$ | 3 |
| Wired link propagation delay | $D_{wd}$ | 0.5 ms |
| Wired link bandwidth | $BW$ | 100 Mbps |
| MCR user velocity | $v$ | [10,30] m/s |
| PU arrival rate | $\lambda$ | [1,6] |
| PU service rate | $\mu_{cp}$ | [0.9,3] |
| MCR user service rate | $\mu_{mcr}$ | [0.6,3] |
| Number of channels in each spectrum band | $N$ | 5 |
| BA radius | $L_{BA}$ | 750 m |
| EA radius | $L_{EA}$ | 1500 m |
| Link layer delay | $t_{L2}$ | 45.35 ms |
| DAD delay | $t_{DAD}$ | 1000 ms |
| Handoff preparation time | $t_{prep}$ | 100 ms |
| RF reconfiguration delay | $t_{rcfg}$ | 300 ms |
| Sensing synchronization time | $t_{syn}^{sen}$ | 25 ms |
| Sensing delay | $t_{sen}$ | 25 ms |
| Spectrum decision delay | $t_{dec}$ | 25 ms |
| Transmission synchronization time | $t_{syn}^{tx}$ | 25 ms |

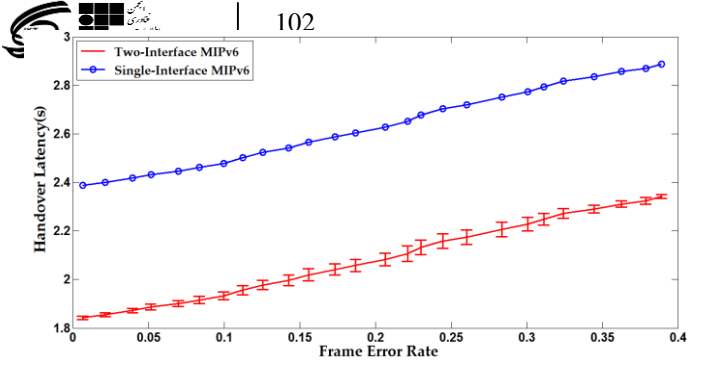

Fig. 5. Intercell/Interpool handoff latency for spectrum mobility scenario versus frame error rate.

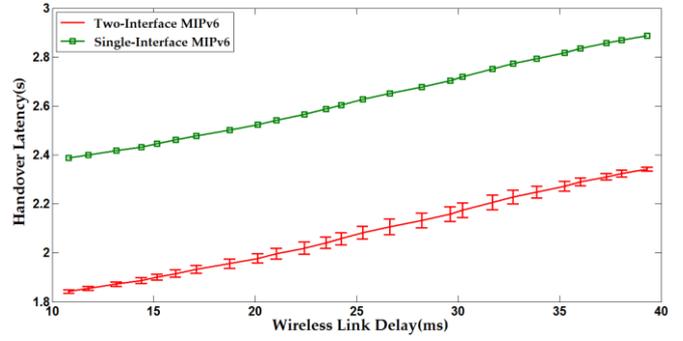

Fig. 6. Intercell/Interpool handoff latency for spectrum mobility scenario versus wireless link delay.

where $t_{prep}$ is the delay to determine handoff type which is called handoff preparation time, $t_{rcfg}$ is the radio interface RF front end reconfiguration latency, $t_{syn}^{sen}$ is the sensing synchronization time, $t_{sen}$ is the sensing operation latency, $t_{dec}$ is the time to determine the proper spectrum band and $t_{syn}^{tx}$ is the transmission synchronization time on the newly selected spectrum band. The term $t_{MD}$ in (14) is related to transmission of Router Solicitation (RS) and Router Advertisement (RA) messages which constitute movement detection latency. The expression for $t_{MD}$ is given as follows:

$$t_{MD} = d_{wl}(L_{RS}) + d_{wl}(L_{RA}). \tag{16}$$

Finally the term $t_{REG}$ in (14) is related to BU latency and return routability procedure delay components. Its expression can be further broken down to:

$$t_{REG} = t_{BU-HA} + t_{BA-HA} + \max(t_{HoTI} + t_{HoT}, t_{CoTI} + t_{CoT}) + t_{BU-CN}, \tag{17}$$

where $t_{BU-HA}$ and $t_{BA-HA}$ are the BU latency sent from MCR user to HA and Binding Acknowledgement (BA) latency sent from HA to MCR user respectively. The expression for binding registration latency to HA is derived as follows:

$$t_{BU-HA} = d_{wl}(L_{BU-HA}) + d_{wd}(L_{BU-HA} + h_{H-A})$$
$$t_{BA-HA} = d_{wl}(L_{BA-HA}) + d_{wd}(L_{BA-HA} + h_{H-A}). \tag{18}$$

The delay components concerned with return routability procedure in (17) are expressed as $t_{HoTI}$, $t_{HoT}$, $t_{CoTI}$ and $t_{CoT}$. These equations are as follows:

$$t_{HoTI} = d_{wl}(L_{HoTI}) + d_{wd}(L_{HoTI}, h_{H-A} + h_{C-H})$$
$$t_{HoT} = d_{wl}(L_{HoT}) + d_{wd}(L_{HoT}, h_{H-A} + h_{C-H}), \tag{19}$$

$$t_{CoTI} = d_{wl}(L_{CoTI}) + d_{wd}(L_{CoTI}, h_{C-A})$$
$$t_{CoT} = d_{wl}(L_{CoT}) + d_{wd}(L_{CoT}, h_{C-A}), \tag{20}$$

where $h_{C-A}$ in equation (23) is $h_{C-G} + h_{G-A}$. Finally the term $t_{BU-CN}$ in (17) is the binding registration latency to the CN which is given by:

$$t_{BU-CN} = d_{wl}(L_{BU-CN}) + d_{wd}(L_{BU-CN}, h_{C-A}) + d_{wl}(L_D) + d_{wd}(L_D, h_{C-A}). \tag{21}$$

The BA message is piggybacked with data packets being sent from the CN to the MCR user.

C. Spectrum Mobility Analysis

As explained in section 3 the Intracell/Intrapool handoff occurs when PU activity is detected in the spectrum and there are free spectrum holes in the current cell. On the other hand the Intercell/Interpool handoff occurs when PU activity is detected in the spectrum without any spectrum opportunity or the cell becomes overloaded. In this section the probability of these types of handoffs across handoff latencies are obtained.

Here the probability of Intracell/Intrapool handoff in the BA, $P_{intra/intra}^{sm\_ba}$ is derived as:

$$P_{intra/intra}^{sm\_ba} = p_{succ}^{sm} P_{on}. \tag{22}$$

The Intracell/Intrapool handoff occurs when the PU appears and the MCR user successfully switches to another free spectrum band. The handoff latency in this case is zero because the secondary radio interface has switched to the free spectrum hole previously.

Whereas the BA of current cell does not have enough spectrum bands, the MCR user has to perform Intercell/Interpool handoff upon PU arrival. Besides when PU appears in the ES, the Intercell/Interpool handoff for MCR users in EA is inevitable because it does not have any available option in that area. Upon PU arrival or capacity overload in the EA the MCR user has to evacuate the current cell. The following is the probability of Intercell/Interpool handoff:

$$P_{inter/inter}^{sm} = P_{over}^i + P_{on}. \tag{23}$$

The secondary radio interface reconfigures its RF front end and performs reassociation, reauthentication, sensing and decision operations while the primary one is involving in data communication. Therefore the MIPv6 handover latency for

Intercell/Interpool handoff scenario, $L^{sm}_{inter/inter}$ (Cases B and C) can be derived as follows:

$$L^{sm}_{inter/inter} = t_{MD} + t_{DAD} + t_{REG}. \qquad (24)$$

V. NUMERICAL ANALYSIS RESULTS

In this section the performance evaluation of the proposed spectrum mobility management technique in terms of handover latency, is carried out and the obtained results are presented and compared with a single radio interface MIPv6 user. The system parameter values used in the numerical analysis are given in Table 2. Other parameters used for handover latency computation are as follows: $h_{C-H} = 4$, $h_{C-G} = 6$, $h_{H-G} = 4$, $h_{G-A} = 4$ and $L_f = 19\ bytes$ [19]. Besides the parameters MCR user velocity $v$, PU arrival rate $\lambda$, PU service rate $\mu_{cp}$ and MCR user service rate $\mu_{mcr}$ are uniformly distributed in the specified spans.

A. Handover Latency

The evaluation of handover latency has been estimated against frame error rate $\sigma_f$ and wireless link delay $D_{wl}$ parameters. The values of $\sigma_f$ and $D_{wl}$ parameters have been uniformly distributed in [0,0.4] and [10,40]ms respectively.

*1) Spectrum Mobility Performance Evaluation*

In this subsection the performance of spectrum mobility has been evaluated in terms of handover latency for Intercell/Interpool handoff in the BA or EA. Various handoff types in the case of spectrum mobility and user mobility management are demonstrated according to frame error rate and wireless link delay parameters. As explained in section 6.4 the MIPv6 handover latency for Intracell/Intrapool handoff is zero for MCR users equipped with two radio interfaces. However the value of MIPv6 Intracell/Intrapool handoff for a single radio interface one would be the same as $t_{SM}$ without $t_{rcfg}$ delay component. Fig. 5 shows the handover latency of Intercell/Interpool handoff against $\sigma_f$ for spectrum mobility scenario. As $\sigma_f$ increases, the probability of erroneous packets becomes large which leads to increasing the mobility signaling retransmission over the wireless link and consequences high handover latency. Fig. 6 presents the handover latency against wireless link delay. An increase in the value of $D_{wl}$, leads to increase in frame delay and results raising the value of handover latency proportionally. With respect to Figs. 5 and 6 it is observed that about 21 percent of reduction in MIPv6 handover latency has been obtained in the proposed two radio interfaces MCR user model compared to the single radio interface one in the case of Intercell/Interpool handover of spectrum mobility scenario.

VI. CONCLUSION

In this paper the integration of the well-known mobility management protocol in the IPv6 networks, MIPv6 and the infrastructure-based CRN paradigm has been considered. Exploiting the multihoming solution in MIPv6 in the context of CRNs has been employed for the performance evaluation of IPv6 mobility management mainly focusing on handover latency as a performance factor. It is assumed that the mobile device in the cognitive radio environment to be equipped with two radio interfaces. While the primary radio interface is communicating with the CN in MIPv6 protocol, the secondary one proactively senses and scans the spectrum bands in its spectrum pool. Before PU appearance using the past spectrum usage history, the secondary radio interface switches to an available free spectrum band and the role of two radio interfaces swaps. Various handoff types have raised according to spectrum mobility management functionalities. The performance evaluation of the proposed two radio interfaces MCR user has been obtained and compared with traditional mobility management protocols in terms of handover latency. Numerical results show that the proposed two radio interfaces model has greatly improved the performance of mobility management.

REFERENCES

[1] B. Sousa, M. Silva, K. Pentikousis, and M. Curado, "A multiple care of addresses model," *ISCC*, pp. 485-490, Jul. 2011.

[2] C. Perkins, D. Johnson, and J. Arkko, "Mobility support in ipv6," *IETF RFC 6275*, Jul. 2011.

[3] S.R. Mugunthan and C. Palanisamy, "A dynamic interoperability mobility management architecture for mobile personal networks," *Wireless Pers Commun*, vol. 83, pp. 1683-1697, Mar. 2015.

[4] A. M. Miyim, M. Ismail, and R. Nordin, "Vertical handover solutions over lte-advanced wireless networks: an overview," *Wireless Pers Commun*, vol. 77, pp. 3051-3079, Mar. 2014.

[5] J.M. Barja, C.T. Calafate, J.C. Cano, and P. Manzoni, "An overview of vertical handover techniques: Algorithms, protocols and tools," *Computer communications*, vol. 34, pp. 985-997, Feb. 2010.

[6] S. Gundavelli, K. Leung, V. Devarapalli, K. Chowdhury, and B. Patil, "Proxy mobile ipv6," *IETF RFC 5213*, Aug. 2008.

[7] R. Koodli, "Mobile ipv6 fast handovers," *IETF RFC 5568*, Jul. 2009.

[8] H. Soliman, C. Castelluccia, K. ElMalki, and L. Bellier, "Hierarchical mobile ipv6 (hmipv6) mobility management," *IETF RFC 5380*, Oct. 2008.

[9] H. Y. Jung, E. A. Kim, J.W. Yi, and H. H. Lee, "A scheme for supporting fast handover in hierarchical mobile ipv6 networks," *J. ETRI*, vol. 27, no. 6, pp. 798-801, Dec. 2005.

[10] I. Al-Surmi, M. Othman, and B. M. Ali, "Mobility management for ip-based next generation mobile networks: review, challenge and perspective," *JNCA*, vol. 35, pp. 295-315, 2012.

[11] I.F. Akyildiz, W.Y. Lee, M.C. Vuran, and S. Mohanty, "Next generation/dynamic spectrum access/cognitive radio wireless networks: a survey," *J. Computer Networks*, vol. 50, pp. 2127-2159, Sep. 2006.

[12] M. J. Kaur, M. Uddin, and H. K. Verma, "Role of cognitive radio on 4g communications a review," *J. Emerging Trends in Computing and Information Sciences*, vol. 3, NO. 2, Feb. 2012.

[13] A. Viziello, I.F. Akyildiz, R. Agustí, L. Favalli, and P. Savazzi, "Cognitive radio resource management exploiting heterogeneous primary users and a radio environment map database," *J. Wirel. Net*. Dec. 2012.

[14] Z. Ping, L. Yang, F.Z. Yong, Z.Q. Xun, L. Qian, and X. Ding, "Intelligent and efficient development of wireless networks: a review of cognitive radio networks," *J. Chinese Sci. Bulletin*, vol. 57, no. 28, pp. 3662-3672, Oct. 2012.

[15] K. Vasu, S. Mahapatra, and C.S. Kumar, "A comprehensive framework for evaluating ipv6 based mobility management protocols," *Wireless Pers Commun*., vol. 78, pp. 943-977, Apr. 2014.

[16] R. Li, J. Li, K. Wu, Y. Xiao, and J. Xie, "An enhanced fast handover with low latency for mobile ipv6," *IEEE Trans. wireless communications*, vol. 7, no. 1, pp. 334-342, Jan. 2008.

[17] C. Huang, M. Chiang, and C. Lin, "A proactive mobile-initiated fast handoff scheme using the multihomed approach", *Wirel. Commun. Mob. Comput.* vol. 9, pp. 1194-1205, 2009.

[18] C.W. Lin, "Fast handoff using the multihomed technique," MSc thesis, Dept. of Computer Science and Information Eng., National Cheng Kung Univ., Jun. 2005.

[19] I.F. Akyildiz, J. Xie, and S. Mohanty, "A survey of mobility management in next-generation all-ip-based wireless systems," *IEEE Wirel. Commun*. vol. 11, pp. 16-28, Aug. 2004.

[20] Y.S. Chen, C.H. Cho, I. You, and H.C. Chao, "A cross-layer protocol of spectrum mobility and handover in cognitive LTE networks," *J. simpat*. vol. 19, pp. 1723-1744, Oct. 2010.

[21] W.Y. Lee and I.F. Akyildiz, "Spectrum-aware mobility management in cognitive radio cellular networks," *IEEE Trans. Mobile Computing*, vol. 11, no. 4, pp. 529-542, Apr. 2012.

[22] Y. Zhang, "Spectrum handoff in cognitive radio networks: opportunistic and negotiated situations," *IEEE International Conference on Communications*, pp. 1-6, Jun. 2009.

[23] J. H. Lee, J. M. Bonnin, I. You, and T. M. Chung, "Comparative handover performance analysis of ipv6 mobility management

protocols," *IEEE Trans. Industrial Electronics*, vol. 60, no. 3, pp. 1077-1088, Mar. 2013.